\documentstyle[12pt,aps]{revtex}

\include{psfig}

\input epsf

\begin{document}
\title{
\hfill\parbox[t]{2in}{\rm\small\baselineskip 14pt
{JLAB-THY-00-10  }\vfill~} \\
\hfill\parbox[t]{2in}{\rm\small\baselineskip 14pt
{(corrected version)}\vfill~}
\vskip 0.5cm
On The Origin of the OZI Rule in QCD \\ 
}

\vskip 0.25cm

\author{Nathan Isgur}
\address{Jefferson Lab, 12000 Jefferson Avenue,
Newport News, Virginia  23606}

\author{H.B. Thacker}
\address{Department of Physics, University of Virginia,
Charlottesville, Virginia 22901}

\maketitle

\vspace{0.5 cm}

\begin{abstract}

		The OZI rule is prominent in hadronic phenomena only
because OZI violation  is typically an order of magnitude
smaller than expected from large $N_c$ arguments.
With its standard $^3P_0$ pair creation operator for hadronic decays by
flux tube breaking, the quark model respects the OZI rule at tree level and
exhibits the cancellations between OZI-violating meson loop diagrams
required for this dramatic suppression. However, if the quark model
explanation for these cancellations is correct, then OZI violation would be
expected to be large in the nonet with
the same quantum numbers as the pair creation operator: the $0^{++}$ mesons.
Experiment is currently unable to identify these mesons, but we report here
on a
lattice QCD calculation which  confirms that the OZI rule arises
from QCD in the vector and axial vector mesons as observed, and finds a
large violation of the rule in the scalar mesons as anticipated by the
quark model. In view of this result, we make some remarks on possible
connections between the
$^3P_0$ pair creation model, scalar mesons,
and the $U_A(1)$ anomaly responsible for the large OZI violation which
drives the $\eta '$ mass. In particular, we note that our result favors the
large
$N_c$ and not the instanton interpretation of the solution to
the $\eta '$ mass problem.

\end{abstract}
\newpage

\section {Background}
\label{sec:background}

\begin{figure}
  \setlength{\unitlength}{0.9 mm}
  \begin{centering}
  \begin{picture}(120,130)(0,0)
    \thicklines

\put(20,20){\line(0, 1){90}}
\put(30,20){\line(0, 1){90}}
\put(35,65){\line(1, 0){1}} \put(40,65){\line(1, 0){1}}
\put(45,65){\line(1, 0){1}}
\put(35,66){\line(1, 0){1}} \put(40,66){\line(1, 0){1}}
\put(45,66){\line(1, 0){1}}
\put(35,65){\line(0, 1){1}} \put(40,65){\line(0, 1){1}}
\put(45,65){\line(0, 1){1}}
\put(36,65){\line(0, 1){1}} \put(41,65){\line(0, 1){1}}
\put(46,65){\line(0, 1){1}}

\put(50,20){\line(0, 1){90}}
\put(60,20){\line(0, 1){90}}

    \put(80,110){\oval(10,80)[bl]} \put(80,110){\oval(10,80)[br]}

\put(74,113){\Large $\bar s$}                \put(84,113){\Large $s$}

  \end{picture}
  \caption[x]{A typical hairpin reaction, where the two lines-ellipsis-two
lines on the left denote an arbitrary
OZI allowed process and an $s \bar s$ hairpin is shown for concreteness.
Note that
gluonic fields and closed $q \bar q$ loops are not represented since the
external quark line topology is all that is relevant to the
rule.}
   \label{fig:OZIreaction}
  \end{centering}
\end{figure}
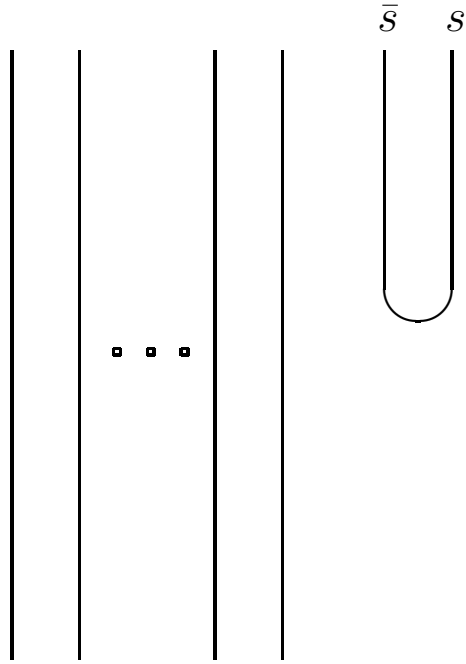

   The phenomena which led to the formulation of the OZI rule
\cite{Zweig,otherOZI}
have had a definitive impact on our understanding of strong interactions.
The fact that  ``aces"
(i.e.,  quarks) led to a simple interpretation of the
properties of the
$\phi$ meson was clearly a very important clue for Zweig \cite{Zweig} since
it was
natural for the
$\phi$  to be pure $s \bar s$
and for certain $\phi$ production cross sections to be small so long as
``hairpin graphs"
were dynamically suppressed (see Fig. \ref{fig:OZIreaction}).

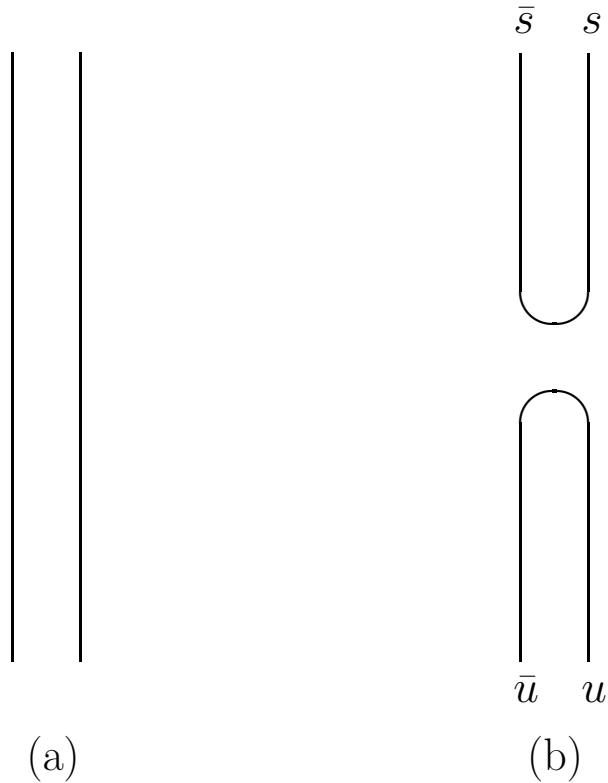
\begin{figure}
  \setlength{\unitlength}{0.9 mm}
  \begin{centering}
  \begin{picture}(120,130)(0,0)
    \thicklines

    \put(20,20){\line(0, 1){90}}
    \put(30,20){\line(0, 1){90}}

    \put(100,20){\oval(10,80)[tl]}  \put(100,20){\oval(10,80)[tr]}
    \put(100,110){\oval(10,80)[bl]} \put(100,110){\oval(10,80)[br]}
    \put(94,14){\Large $\bar u$}                \put(104,14){\Large $u$}
    \put(94,113){\Large $\bar s$}        \put(104,113){\Large $s$}

    \put(22,4){\Large (a)}     \put(96,4){\Large (b)}

  \end{picture}
\bigskip\bigskip
  \caption[x]{Quark line diagrams associated with the OZI 
rule in meson nonets: (a) a normal
OZI-conserving quark-antiquark ``scattering" process, (b) a typical
double-hairpin ``annihilation"
process leading to flavor-mixing in meson wave functions.
Since the OZI-violating reactions of Fig. \ref{fig:OZIreaction} could occur
through OZI-allowed meson emission followed by  flavor mixing,
these
processes are the simplest manifestations of OZI violation. Note that gluonic
fields and closed $q \bar q$ loops are not represented since the external quark
line topology is all that is relevant to the rule.   }
   \label{fig:OZI}
  \end{centering}
\end{figure}

	  The dynamics behind the suppression of hairpin graphs in QCD has
remained unexplained.
The {\it phenomenology} of meson mixing angles in QCD-based
quark models was described in the mid-1970's in a number of papers
\cite{DGG,NImix,FritzschMinkowski}.  In such models, processes with the
quark line topology of the
double-hairpin graphs of Fig. \ref{fig:OZI}(b) (but with arbitrary time
orderings) modify
the  quark-antiquark transition amplitudes from the totally flavor diagonal
form associated with the ``scattering" quark line topology of Fig.
\ref{fig:OZI}(a), namely
\begin{equation}
\bf T =
\left[ \matrix{S & 0 & 0 & 0 \cr
0 & S & 0 & 0 \cr
0 & 0 & S & 0 \cr
0 & 0 & 0 & S \cr
} \right] ~~,
\end{equation}
(for illustrative purposes we have suppressed all space-time labels and
specialized to the
case of $SU(2)$ flavor where the matrix spans
the basis $u \bar d$, $d \bar u$, $u \bar u$, $d \bar
d$) by the addition of the annihilation amplitudes $A$
\begin{equation}
{\bf \Delta T} =
\left[ \matrix{0 & 0 & 0 & 0 \cr
0 & 0 & 0 & 0 \cr
0 & 0 & A & A \cr
0 & 0 & A & A \cr
} \right] ~~.
\end{equation}
Using this framework \cite{subtlety}, it was noted that the OZI mixing
amplitude $A$
characterizing
Fig. \ref{fig:OZI}(b) was of order 10 MeV in the established meson nonets,
with the sole exception
of the ground state pseudoscalar meson nonet, where $A$ is  an order of
magnitude larger.  These
observations were consistent with the pattern one would expect for heavy
quarkonia where the ground
state pseudoscalar double-hairpin is larger than the vector double-hairpin by
one factor of $(\alpha_s/\pi)^{-1}$,
and excited state double-hairpins are suppressed by having vanishing wave
functions at $\vec r=0$.
However, an explanation for this pattern in light quark systems was lacking.

\bigskip\bigskip

\vspace{0.3cm}

     The large size of the ground state pseudoscalar double-hairpin is a
manifestation of  the  ``$U_A(1)$ problem" \cite{UA(1)Problem}:  the
equations of motion of QCD, taken naively, would imply that
spontaneous chiral symmetry breaking leads
to {\it nine} and not just eight Goldstone bosons \cite{NambuGoldstone},
but the large mass of the $\eta'$ seems to disqualify it from the role of
the flavor singlet Goldstone boson.
However, the $U_A (1)$
current is anomalous, and by the late 1970's it was understood through the
study of instantons
\cite{instantons,tHooftInstantons} that the anomaly leads to a
nonconservation of the $U_A (1)$
charge and thereby to the evasion of Goldstone's theorem in the flavor
singlet channel when chiral symmetry is spontaneously
broken.  The  connection between the quark model picture of double-hairpins
and instantons was discussed
by Witten \cite{WittenUA(1)}, Veneziano \cite{Veneziano}, and others, who
explored
more generally the conflict between instantons and
the large $N_c$ expansion \cite{LargeNc}.

\bigskip

\vspace{0.3cm}

\bigskip\bigskip

     (The reader familiar with instanton lore may be puzzled by the
connection between the
annihilation amplitudes  $A_{OZI}^{0^{-+}}$ of Eq. (2) in the pseudoscalar
mesons and
instanton-induced effects in the pseudoscalar mesons.  The latter effects
are associated with the
't Hooft interaction
\cite{tHooftInstantons} which (in our illustrative $SU(2)$ flavor case)
leads to $u \bar
u \rightarrow d \bar d$ and
$d \bar d \rightarrow u \bar u$ but {\it not} the diagonal entries in Eq.
(2) for $\Delta T$
corresponding to $u \bar u \rightarrow u \bar u$ or $d \bar d \rightarrow d
\bar d$ transitions.
Recall, however, that the 't Hooft interaction also has
$u \bar d \rightarrow u \bar d$ and $d \bar u \rightarrow d \bar u$
interactions, i.e., the $S$-like amplitudes of Eq. (1).  Thus the
instanton-induced interactions
also admit the decomposition of Eqs. (1) and (2) with $S= -A$.  We
will
elaborate upon this point below.)

\bigskip\bigskip

\begin{table}
  \caption {OZI-violating amplitudes in meson nonets. These amplitudes are
defined
to be the contribution of the $u \bar u \rightarrow d \bar d$ double-hairpin
to the nonet
mass matrix.}
\vspace{0.1cm}
 \begin{center}
  \begin{tabular}{|ccccc|}
& nonet  & empirical  & quark model loop & \\
&    &  $A_{OZI}^{J^{PC}}$ (MeV) & contribution to $A_{OZI}^{J^{PC}}$
(MeV)$^\dagger$  & \\
\hline
&$ 0^{-+}$ & $+400 \pm 200~^*$  & - - -  &  \\
&$1^{--}$ & $+7 \pm 1$  & $-2 \pm 4$  &  \\
&$2^{++}$ & $-22 \pm 3$  & $+6 \pm 14$  &  \\
&$1^{++}$ & $+11 \pm 15$  & $+12 \pm 12$  &  \\
&$0^{++}$ & see text  & $-450 \pm 200~^*$  &  \\
&$1^{+-}$ & $-32 \pm 12$  & $-15 \pm 7$  &  \\
&$3^{--}$ & $-12 \pm 4$  & $+4 \pm 7$  &  \\
&$4^{++}$ & $+6 \pm 18$  & $+16 \pm 7$  &  \\
  \end{tabular}
  \label{tab:OZI}
 \end{center}

$^*$ See Ref. \cite{mass2}.

\medskip

$^\dagger$The quoted ``theoretical
error" assigned here is the range quoted in Ref. \cite{GIonOZI} for meson
loop processes from
reasonable parameter variations.

\end{table}

     The large $N_c$ expansion is the only known field-theoretic basis for
the general success
of the valence quark model, Regge phenomenology, the observed narrowness of
resonances, and the OZI
rule.  In particular, the  OZI-violating meson mixing amplitudes of Fig.
\ref{fig:OZI}(b) are all
of order
$1/N_c$.  Ironically, such a  suppression of these amplitudes seems
perfectly consistent with the
effects in the pseudoscalar mesons, but not strong enough to account for
the extremely small
amplitudes seen in other nonets. See the second column of Table
\ref{tab:OZI}.

\begin{figure}
  \setlength{\unitlength}{0.9 mm}
  \begin{centering}
  \begin{picture}(120,130)(0,0)
    \thicklines
    \put(50,10){\oval(10,40)[tr]}  \put(70,10){\oval(10,40)[tl]}
    \put(50,50){\oval(10,40)[bl]}  \put(70,50){\oval(10,40)[br]}
    \put(60,50){\oval(10,30)[bl]} \put(60,50){\oval(10,30)[br]}
    \put(50,100){\oval(10,40)[br]} \put(70,100){\oval(10,40)[bl]}
    \put(50,60){\oval(10,40)[tl]}  \put(70,60){\oval(10,40)[tr]}
    \put(60,60){\oval(10,30)[tl]} \put(60,60){\oval(10,30)[tr]}
    \put(45,60){\line(1,-1){10}}   \put(75,60){\line(-1,-1){10}}
    \put(45,50){\line(1, 1){4}}    \put(55,60){\line(-1,-1){4}}
    \put(65,60){\line(1,-1){4}}   \put(75,50){\line(-1,1){4}}
    \put(54,4){\Large $\bar u$}                \put(64,4){\Large $u$}
    \put(54,103){\Large $s$}        \put(64,103){\Large $\bar s$}

  \end{picture}
  \caption[x]{Two sequential OZI-allowed processes can lead to the topology
of Fig.
\ref{fig:OZI}(b). }
   \label{fig:OZIloop}
  \end{centering}
\end{figure}
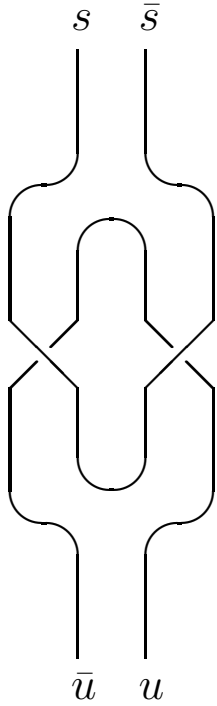

     The unexpected suppression of most OZI-violating amplitudes
beyond a simple factor of $1/N_c$ is elevated from a
dynamical puzzle to  a paradox
when the various time-orderings of Fig. \ref{fig:OZI}(b) are projected into a
hadronic basis. In such a basis, flavor mixing could arise through an
intermediate
glueball, through an instantaneous interaction, or via a hadronic loop
process in
which Fig. \ref{fig:OZI}(b) has the time-ordering shown in Fig.
\ref{fig:OZIloop}.
The paradox arises from the observation \cite{LipkinOZI}
that these  OZI-violating hadronic loop processes can proceed by
sequential {\it OZI-allowed}
vertices with known and unsuppressed strengths.  These  hadronic loop
diagrams may
be associated with contributions to meson
propagators arising from second order (real and virtual) decay processes,
and as such are of
order $(1/\sqrt{N_c})^2$, as expected.  This factor of $1/N_c$ is also
perfectly consistent with
the observation that  the {\it imaginary} parts of these propagators
give the $1/N_c$-suppressed meson widths which are generally of order of
{\it hundreds} of MeV.
Nevertheless, OZI phenomenology requires that the  $1/N_c$-suppressed
{\it real} parts (from the full meson spectrum and not just the
kinematically allowed part) be an order of magnitude smaller.  Explicit
model calculations substantiate the generic result
that individual hadronic channels of the type depicted in
Fig. \ref{fig:OZIloop}
would  indeed contribute hundreds of MeV to OZI-violating meson mixing.
Thus even if the other possible sources of OZI violation (from the other
time-orderings) were dynamically suppressed, these hadronic loop diagrams
would seem
to spoil the OZI rule. This rule requires that $A_{OZI} <<m_s-m_d$ so that
the nonet
and not the $SU(3)$ limit is realized. Thus it is a necessary (but not
sufficient)
condition for the OZI rule that there be some conspiracy between hadronic loop
processes which suppresses them below their expected $1/N_c$ strength
\cite{caveat}.

\bigskip

\section {A Proposed Resolution}
\label{sec:resolution}

   The authors of Ref. \cite{GIonOZI} proposed a resolution of this
paradox.  They
examined the OZI-violating amplitudes $A_{OZI}$ in non-pseudoscalar
channels from the
{\it complete} tower of hadronic loop processes to determine if
``miraculous" cancellations
between the hundred-MeV-scale real parts
of individual channels could be responsible for the suppression
of the sum over channels
beyond a simple power of $1/N_c$. To make such a
calculation one must
have a complete model for  meson trilinear vertices since, if such a
conspiracy is to occur, it
will have to be based on an underlying pattern of coupling {\it strengths}
and {\it signs}.  The
nonrelativistic quark model is complete in this sense:  using the standard
$^3P_0$ pair creation operator for hadronic decays by flux tube
breaking \cite{3P0} and valence quark model wave functions, all trilinear
vertices and their associated form factors are prescribed.  Since the model
is a
nonrelativistic one, only the time-ordering of Fig. \ref{fig:OZIloop} with
a two meson intermediate
state can be calculated in this way, but within this framework Ref.
\cite{GIonOZI} shows that in
general a ``miraculous" cancellation between channels does indeed occur.
This cancellation occurs
between groups of intermediate meson states that might have been difficult
to anticipate {\it a
priori}. Consider the prototypical case of $\omega-\phi$ mixing where
$\omega \rightarrow (AB)_L \rightarrow \phi$ with $L$ the $AB$ relative
angular momentum.  The intermediate states contributing to
Fig. \ref{fig:OZIloop} are
$(K \bar K)_P$,
$(K \bar K^*)_P$,
$(K^* \bar K^*)_P$,
$(K^* \bar K^*_0)_S$,
$(K \bar K_{a_1})_S$,
$(K \bar K_{a_1})_D$,
$(K^* \bar K_{a_1})_S$,
$(K^* \bar K_{a_1})_D$,
$(K \bar K_{b_1})_S$,
$(K \bar K_{b_1})_D$,
$(K^* \bar K_{b_1})_S$,
$(K^* \bar K_{b_1})_D$,
$(K \bar K^*_2)_D$,
$(K^* \bar K^*_2)_S$,
$(K^* \bar K^*_2)_S$, ...
where $K$ and $K^*$ are the ground state ($\ell=0$) pseudoscalar and vector
mesons, and
$K^*_0$, $K_{a_1}$, $K_{b_1}$, and $K^*_2$  are
the first excited state ($\ell=1$) strange mesons with $J^P=0^+, 1^+, 1^+$,
and $2^+$
which would be
associated with the $a_0$, $a_1$, $b_1$, and $a_2$ octets in the $SU(3)$
limit.  (Note that the
ellipsis denotes more highly excited intermediate states, including ones in
which {\it each} leg
of the intermediate state is excited, and that charge conjugate
intermediate states are implied.)
As expected on the basis of the previously described arguments, a typical
channel in this sum
contributes of order 100 MeV to $A_{OZI}^{1^{--}}$. However, intermediate
states with the same {\it
total orbital angular momentum} but opposite  values of $(-1)^L$ tend to
cancel.  Thus, for example,
the
$(\ell_A=0, \ell_B=0)_P$
channels with
$L_{total} \equiv \ell_A+\ell_B+L=1$  all have the same sign, but they
strongly cancel against the $(\ell_A=0, \ell_B=1)_S$+$(\ell_A=1, \ell_B=0)_S$
channels!

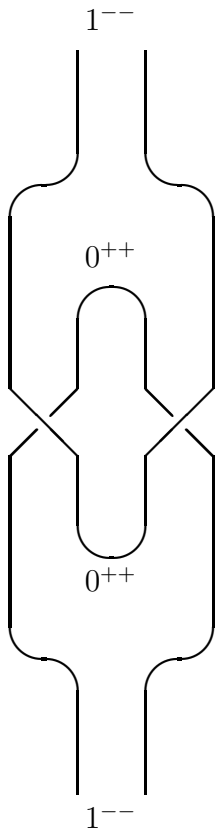
\begin{figure}
  \setlength{\unitlength}{0.9 mm}
  \begin{centering}
  \begin{picture}(120,130)(0,0)
    \thicklines
    \put(50,10){\oval(10,40)[tr]}  \put(70,10){\oval(10,40)[tl]}
    \put(50,60){\oval(10,60)[bl]}  \put(70,60){\oval(10,60)[br]}
    \put(60,60){\oval(10,30)[bl]} \put(60,60){\oval(10,30)[br]}
    \put(50,120){\oval(10,40)[br]} \put(70,120){\oval(10,40)[bl]}
    \put(50,70){\oval(10,60)[tl]}  \put(70,70){\oval(10,60)[tr]}
    \put(60,70){\oval(10,30)[tl]} \put(60,70){\oval(10,30)[tr]}
    \put(45,70){\line(1,-1){10}}   \put(75,70){\line(-1,-1){10}}
    \put(45,60){\line(1, 1){4}}    \put(55,70){\line(-1,-1){4}}
    \put(65,70){\line(1,-1){4}}   \put(75,60){\line(-1,1){4}}
    \put(56,5){$1^{--}$}
    \put(56,123){$1^{--}$}
    \put(56,40){$0^{++}$}
    \put(56,88){$0^{++}$}

  \end{picture}

\bigskip\bigskip

  \caption[x]{ A graphical representation of the ``spectator plus closure
limit":  in this limit the
$0^{++}$ pair creation operator cannot destroy or create a $1^{--}$ pair. }
   \label{fig:closure}
  \end{centering}
\end{figure}

     The calculation is formidable.  With standard quark model parameters
the form factors are
quite hard and complete convergence is achieved only after summing of order
10 thousand
channels, corresponding to $L_{total} \simeq 10$.  With reasonable
variations of
standard parameters the  contribution of an individual channel waxes and
wanes,
as does the speed
of convergence. However, {\it the underlying mechanism of the
cancellation is simple
and very robust:  $A_{OZI}^{1^{--}}$
is much smaller than its component pieces because of an
approximate ``spectator plus closure limit"}.  This limit is illustrated in
Fig. \ref{fig:closure},
which shows the standard  $^3P_0$ operator with $ J^{PC}=0^{++}$ trying to
create and then
annihilate quark-antiquark pairs with $ J^{PC}=1^{--}$.  If a single two
meson intermediate state
is inserted into this diagram, it will project out pieces of this amplitude
of order $1/N_c$ as
expected, but if the original (final)
$q \bar q$ pair does not distort the $ J^{PC}$ of the produced
(annihilated) pair (the
spectator approximation) a complete set of intermediate states with a
common energy denominator
(the closure approximation) {\it will give zero amplitude}.  Ref.
\cite{GIonOZI} shows that
deviations from this ``spectator plus closure limit" are naturally
small, leading to the observed order
of magnitude suppression of the loop
contribution to $A_{OZI}^{1^{--}}$
relative to $1/N_c$ expectations. See Table \ref{tab:OZI}. The interested
reader
is referred to  Ref. \cite{GIonOZI} for a detailed explanation of the
resiliency of this limit.
This quark model solution to the ``second order paradox" associated with
the OZI
rule also appears to justify the conspiracies
between Regge trajectories required to explain the suppression of cross
sections requiring
``exotic" exchanges (e.g., those with isospin 2) \cite{ReggeExotics}. Since
``exotic"
exchanges can occur by double Regge exchanges (analogous to the second
order loop processes), only
a conspiracy between exchanges (analogous to the conspiracy between loops)
can give the observed
suppression of such cross sections.

     While the order of magnitude suppression of the loop
contribution to $A_{OZI}^{1^{--}}$ is robust, the contribution of
individual channels and the residue after the cancellations have occurred is
model sensitive, so a prediction for the actual value of this amplitude
cannot be made.  This is not a great loss, however, since the accuracy of the
model is very suspect:  its dynamics is
nonrelativistic, and it  has ignored the $Z$-graph  time orderings of Fig.
\ref{fig:OZIloop}. More significantly, any such amplitude would need to be
added to the unknown
pure glue and instantaneous contributions to the $q \bar q
\rightarrow q' \bar q'$
transition before being compared to experiment.  Thus the important
conclusion of
Ref.
\cite{GIonOZI} is the {\it qualitative} one that  the
``second order paradox" can be evaded.

     Ref. \cite{GIonOZI}  confirms that $A_{OZI}$ from meson loop diagrams
is small in not
only the vector  mesons but in all other well-established nonets:  those with
$ J^{PC}=2^{++}, 1^{++}, 1^{+-}, 3^{--}$, and $4^{++} $.  The key, of
course, is that the  nonet $ J^{PC}$ must differ from that of the
$^3P_0$ pair creation operator \cite{missingpseudoscalar}.  From
this simple requirement follows a rather spectacular prediction:  {\it OZI
violation
should be very
strong in the scalar meson nonet}.

     The scalar mesons, and especially the isoscalar scalar mesons which
would display the
effects of OZI violation, have been notoriously difficult to understand
experimentally.  Over
the last thirty years the mass of the lightest isoscalar scalar
meson quoted by the
Particle Data Group has varied between 400 and 1400 MeV, while the quoted
width has varied between and 100 and 1000 MeV.
(We have removed the $f_0(980)$ from this compilation under the presumption
that it is a $K \bar K$
molecule, or this spread of values would be even wider.)  The experimental
status of the
scalar meson nonet becomes even more obscure when one recalls that the
lightest glueball is
expected to have $ J^{PC}=0^{++}$ and a mass around 1.5 GeV.  One can only
say with confidence that
the  experimental situation does not exclude that $A_{OZI}^{0^{++}}$ is large.

     Fortunately, there is an alternative to checking this prediction of the
quark model mechanism against experiment.  We can check it against
calculations from
lattice QCD.

\bigskip

\bigskip

\section {OZI on the Lattice}
\label{sec:lattice}

\subsection {OZI Violation in the Quenched Approximation}

     Matrix elements of the type
$\langle 0 \vert T[\bar q'(y) \Gamma^{J^{PC}} q'(y) ~ \bar q(x)
\Gamma^{J^{PC}} q(x)] \vert 0
\rangle$, where $\Gamma^{J^{PC}}$ carries space-time  indices which
determine the $ J^{PC}$ of the
propagator being studied and $q' \neq q$, describe OZI violation in mesons.
In leading order in
$1/N_c$, such processes can proceed through diagrams of the type depicted
in Fig.
\ref{fig:OZI}(b) (of which Fig. \ref{fig:OZIloop} is the particular
time-ordering relevant to the hadronic loop diagrams), i.e., they
receive leading contributions
in the quenched approximation in which internal quark-antiquark loops are
ignored.  (Of course the
{\it accuracy} of the quenched approximation can be questioned, but this is
irrelevant to the main points of
this paper, including checking a prediction of the quark model in which
internal
quark loops are also neglected.)

     In the absence of OZI violation, the $\omega$-like
${1 \over \sqrt 2}(u \bar u +d \bar d)$ and
$\phi$-like
$s \bar s$ sectors are  segregated and each develops its own tower of meson
excited states of each
allowed
$J^{PC}$. If the OZI-violating amplitudes $A_{OZI}^{J^{PC}}$ in that
channel are small, then in
leading order they simply shift the masses of each state by $ \sim
A_{OZI}^{J^{PC}}$
and create
$\omega-\phi$-like mixing with a mixing angle $ \sim
A_{OZI}^{J^{PC}}/\Delta m$ where
$\Delta m$ is the
unperturbed mass difference between the $\omega$- and $\phi$-like states being
mixed.  In such circumstances the empirical value of $A_{OZI}^{J^{PC}}$ may
be extracted from
either the $\omega-\rho$-like mass difference or the $\omega-\phi$-like
mixing angle and compared
directly with the quenched lattice amplitudes since the latter may be
construed as correctly
representing OZI-violation in the quenched approximation in lowest order
perturbation theory in $A_{OZI}^{J^{PC}}$.

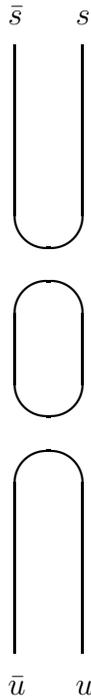
\begin{figure}
  \setlength{\unitlength}{0.9 mm}
  \begin{centering}
  \begin{picture}(120,130)(0,0)
    \thicklines

    \put(60,10){\oval(10,60)[tl]}  \put(60,10){\oval(10,60)[tr]}
    \put(60,100){\oval(10,60)[bl]} \put(60,100){\oval(10,60)[br]}

      \put(60,55){\oval(10,20)[tl]}  \put(60,55){\oval(10,20)[tr]}
      \put(60,55){\oval(10,20)[bl]} \put(60,55){\oval(10,20)[br]}

    \put(54,4){$\Large \bar u$}                \put(64,4){$\Large u$}
    \put(54,103){$\Large \bar s$}        \put(64,103){$\Large s$}

  \end{picture}

\bigskip\bigskip\bigskip

  \caption[x]{A contribution to OZI-violating meson mixing which is
superficially
inconsistent with the quenched approximation. }
   \label{fig:OZIquenched}
  \end{centering}
\end{figure}

     If $A_{OZI}^{J^{PC}}$ is strong, as in the pseudoscalar channel, the
situation is
more  complicated.  In such circumstances two new effects come into play:
the masses of
$\omega$- and $\phi$-like states can be shifted strongly, so that their
mixing angle may not be
determined by their unperturbed mass difference, and treating the mixed
propagator from
$q \bar q \rightarrow q' \bar q'$ in lowest
order in $A_{OZI}^{J^{PC}}$ may not be valid.  The former effect is
straightforward, but the
latter can be complex. For example, a higher
order treatment of $A_{OZI}^{J^{PC}}$ appears to be inconsistent with the
quenched approximation, as shown in Fig.
\ref{fig:OZIquenched}. However, the process depicted in Fig.
\ref{fig:OZIquenched} is one of a
series of processes with internal quark loops which arise from repeated
iteration of the quenched amplitude.
Their effect and that of the diagonal mass shifts
is to create a propagator {\it matrix} with entries corresponding to the
quenched
approximation;  when diagonalized perturbatively this matrix gives the
masses and mixing angles
for weak OZI violation, but for strong OZI violation it may be diagonalized
exactly, thereby
summing the series of sequential applications of $A_{OZI}^{J^{PC}}$.
Another  closely related  possible complication is that a large
$A_{OZI}^{J^{PC}}$ can create strong mixing with the glueball sector,
requiring that the propagator matrix be enlarged yet further.

\bigskip\bigskip\bigskip

\vspace{0.3cm}

\bigskip\bigskip

     For the pseudoscalar mesons, the preceeding discussion of the effects of a
large $A_{OZI}$ are particularly significant.  In the chiral limit with
$A_{OZI}^{ 0^{-+}}=0$, the $U_A(1)$ meson - - -  the $\eta'$ - - -  is also
massless.  As a result,
the quenched OZI-violating amplitude of Fig. \ref{fig:OZIloop} will give
$A_{OZI}^{0^{-+}}$
sandwiched between two massless propagators, i.e., it will give an $\eta'$
propagator that looks
nothing like that of a massive,
$SU(3)$-flavor-mixed $\eta'$.  In this case, to even {\it qualitatively}
relate
the quenched amplitudes to nature one must extract
 $A_{OZI}^{0^{-+}}$ from them and add these amplitudes to the propagator {\it
matrix}
(the broken $SU(3)$ analog of Eq. (1)) which one
diagonalizes exactly.  The resulting full propagator will have a massive
$SU(3)$-flavor-mixed $\eta'$ which sums the single particle effects of
$A_{OZI}^{0^{-+}}$ to all
orders. Further numerical support for this interpretation of the quenched
pseudoscalar double-hairpin comes from the shape of the double-hairpin
propagator as a function of Euclidean time. As discussed below (see also Ref.
\cite{lat98}), this time dependence can be fit very well to the functional
form $ (1+m_{\pi}t)\exp(-m_{\pi}t)$  expected from a mass insertion
vertex surrounded by two propagators of mass $m_{\pi}$.

\subsection {Methods}

The ability to study the double-hairpin diagrams relevant to the OZI rule has
been greatly
improved by two recent developments in lattice QCD methodology. The global
source technique (which we  refer to as the ``allsource'' method)
was introduced several years ago for the purpose of studying
the $\eta'$ mass and the $U_A(1)$ anomaly\cite{Kuramashi}.
In this method, the quark propagator is
calculated from a sum of identical unit color-spin sources located at all
space-time points on the lattice. If this allsource propagator is contracted
over color indices at a {\it given} site, the result is a gauge invariant term
corresponding to a closed quark loop  originating from that site, plus a very
large number of gauge-dependent open loops. The latter terms tend to cancel
due to their random phases, allowing a  determination of
closed loop averages and loop-loop correlators (double-hairpins). The other
recently developed technique which has greatly improved the accuracy of the
results for both double-hairpin calculations and for other chiral studies with
Wilson-Dirac fermions
is the Modified Quenched Approximation (MQA) \cite{lat98}, which provides a
practical resolution of the exceptional configuration problem that has long
plagued such calculations. This method identifies the source of the exceptional
configuration problem as the presence, in some gauge configurations, of exactly
real eigenmodes which are displaced into the physical mass region by the
artificial chiral symmetry breaking  associated with the lattice Wilson-Dirac
operator. By systematically identifying these real eigenmodes and calculating
their contribution to the quark propagators, the corresponding propagator poles
can be extracted and moved to zero quark mass (where they belong). This MQA
procedure has been applied to both the allsource propagators for
double-hairpin calculations as well as to valence quark propagators. The
resulting MQA-improved
propagators have recently been used in an extensive study of quenched
chiral logs
and their relation to the $\eta'$ mass and the $U_A(1)$ anomaly
\cite{chlogs}. As a part of this study, the size and
time-dependence of the pseudoscalar $\eta'$ double-hairpin diagram was
calculated, using the allsource method. Since the pole-shifting procedure has
already been applied to the quark propagators, it requires very little
additional effort to investigate the  vector, axial-vector, and scalar
double-hairpins which determine the spin-parity pattern of OZI mixing. The
results we present here are from a set of 300 quenched gauge configurations
on a $12^3\times 24$ lattice at $\beta=5.7$. In the study of Refs.
\cite{lat98,chlogs}, both naive Wilson and clover improved quark actions
were studied. It was found that, at $\beta=5.7$ with the Wilson action,
substantial lattice spacing effects suppressed the pseudoscalar
double-hairpin, giving a smaller than expected value of ${\bf
A}_{OZI}^{0^{-+}}=$(0.27 GeV)$^2$ for the double-hairpin contribution to the
$\eta'$ mass versus the value (0.49 GeV)$^2$ extracted from
weak-$SU(3)$-breaking mass formulas \cite{mass2}. A much more satisfactory
result
is  obtained from the clover improved quark action. With a clover coefficient
$C_{sw}=1.57$, the pseudoscalar double-hairpin 
gives ${\bf A}_{OZI}^{0^{-+}}=$(0.41 GeV)$^2$.  For the
calculation of OZI-violating amplitudes, we will therefore use the clover
improved quark action only; we also use the
physical charmonium 1S-1P splitting to set the scale
($a^{-1}=1.18$ GeV) for $\beta=5.7$  when we 
quote lattice results in physical units \cite{scales}.

\subsection {Results}

Using the method described in the previous Section, we have calculated the
double-hairpin
contribution to matrix elements of the form
\begin{equation}
\langle\bar{q}'(y)\Gamma^iq'(y)~\bar{q}(x)\Gamma^iq(x)\rangle
\end{equation}
with Hermitian operators generated by the choices
$\Gamma^i=i \gamma_5$ (pseudoscalar),
$\Gamma^i=\gamma^{\mu}$, $\mu=1,2,3$ (vector), 
$\Gamma^i=\gamma^{\mu}\gamma_5$, $\mu=1,2,3$
(axial vector) and $\Gamma^i=1$ (scalar) (the antisymmetric tensor $\sigma^{\mu
\nu}$ does not explore new states: it also has axial vector quantum
numbers). As in standard hadron spectroscopy, we Fourier transform the
space-time propagator
over 3-dimensional time slices at zero 3-momentum and study its
time-dependence. A
particular advantage of the allsource method is that the Fourier transforms
can be
performed over both ends of the meson propagator, unlike the usual case of
a fixed
local source where only one end can be transformed. This provides an
improvement in
statistics which is quite important for the success of the
method. For
the scalar double-hairpin matrix element, the expectation value of a single
scalar loop
is nonzero, and so a constant proportional to $
\langle 0 \vert \bar q q \vert 0 \rangle ^2 $ must be  subtracted from the
above matrix element to get the true correlator.

\vskip 1cm

\begin{table}
  \caption {Quenched lattice OZI-violating amplitudes.}
\vspace{0.1cm}
 \begin{center}
  \begin{tabular}{|ccccc|}
& nonet  & ${\bf A}_{OZI}^{J^{PC}}$ (MeV)$^2$  & $A_{OZI}^{J^{PC}}$
(MeV)$^*$ & \\
\hline
&$ 0^{-+}$ & $+(407\pm 11)^2$     &  $\simeq +290 $   &  \\
&$1^{--}$  & $<(220)^2$              &  $<30$             &  \\
&$1^{++}$  & $<(380)^2$              &  $<60$             &  \\
&$0^{++}$  & $-(1350\pm 90)^2$              &  $\simeq -520$    &  \\
  \end{tabular}
  \label{tab:latticeOZI}
 \end{center}

\medskip

$^*$ For the conversion from ${\bf A}_{OZI}^{J^{PC}}$ extracted from the
lattice
via Eq. (\ref{eq:fit}) to $A_{OZI}^{J^{PC}}$ for comparison to the
amplitudes quoted in Table
\ref{tab:OZI} based
on mass matrices, see  Ref.
\cite{mass2}.

\end{table}

\begin{figure}
\begin{center}
\hskip -0.5cm
\epsfxsize=4.2in 
\epsfbox{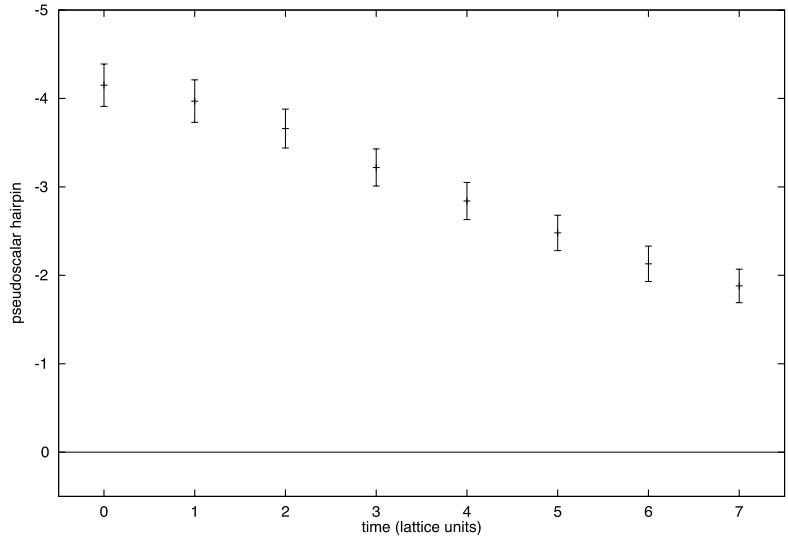}
\end{center}
\caption{The pseudoscalar double-hairpin correlator. Note the scale: this
amplitude is negative.}
\label{fig:psprop}
\end{figure}

\begin{figure}
\begin{center}
\hskip -0.5cm
\epsfxsize=4.2in 
\epsfbox{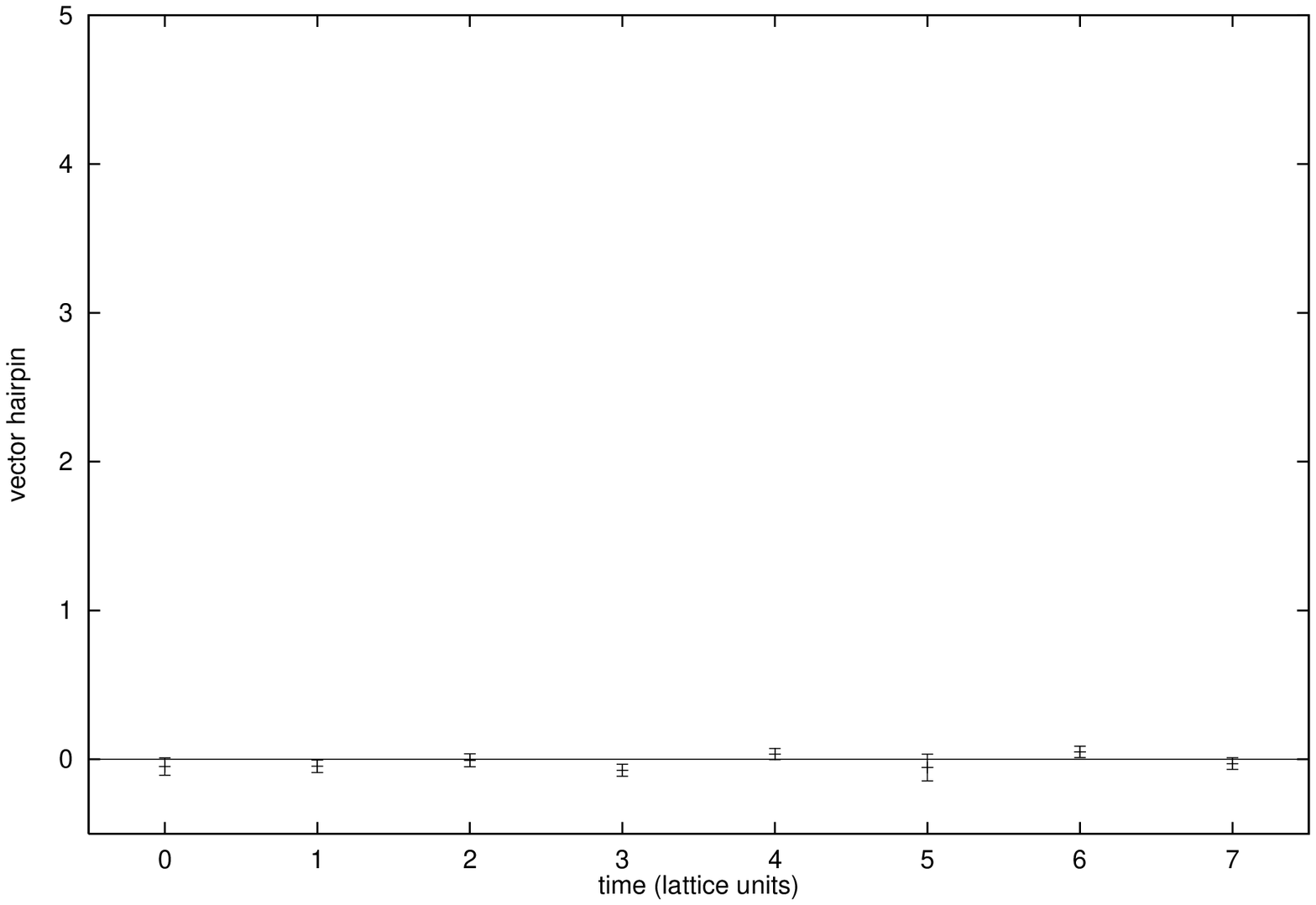}
\end{center}
\caption{The vector double-hairpin correlator.}
\label{fig:vecprop}
\end{figure}

\begin{figure}
\begin{center}
\hskip -0.5cm
\epsfxsize=4.2in 
\epsfbox{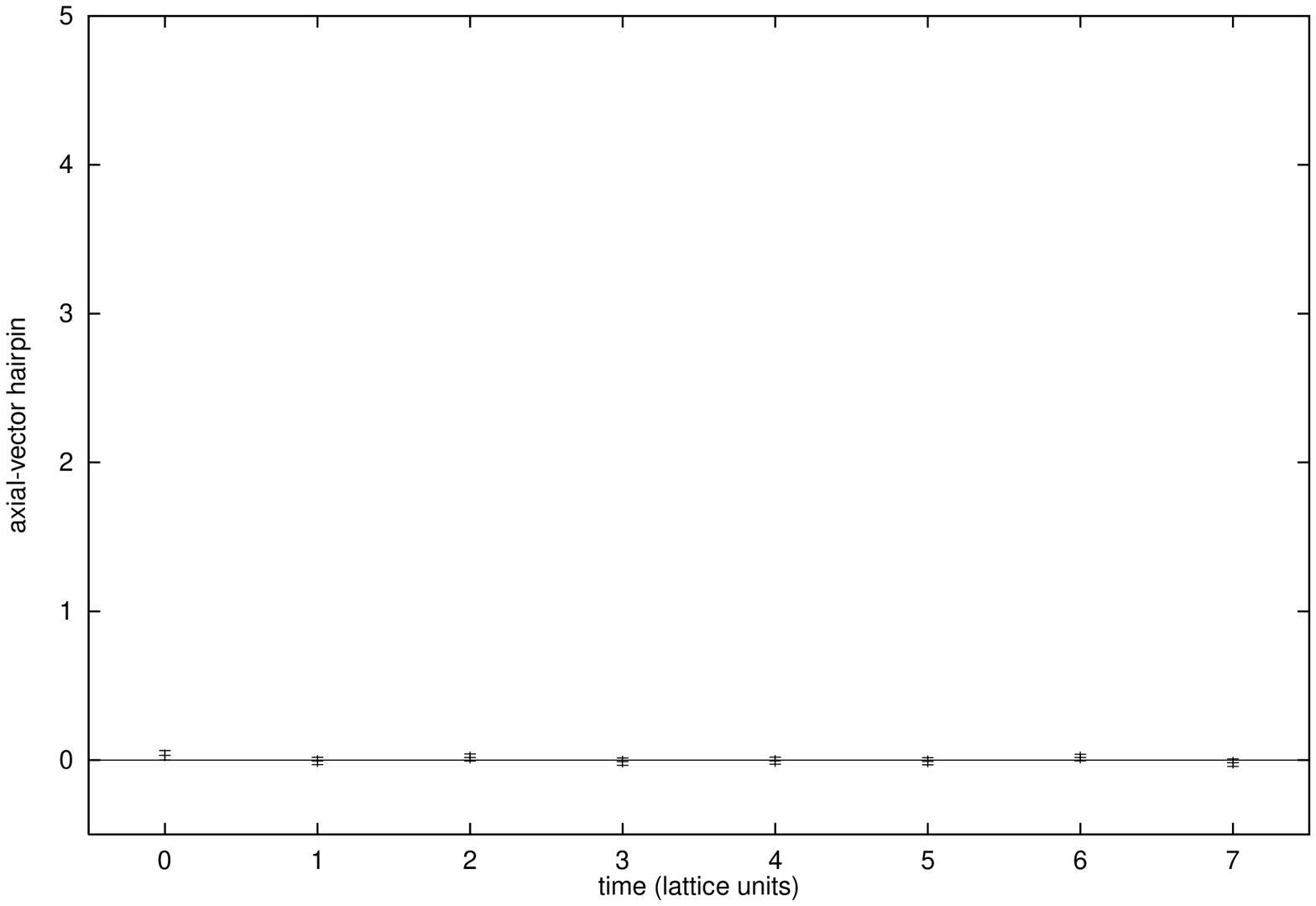}
\end{center}
\caption{The axial vector double-hairpin correlator.}
\label{fig:avecprop}
\end{figure}

\begin{figure}
\begin{center}
\hskip -0.5cm
\epsfxsize=4.2in 
\epsfbox{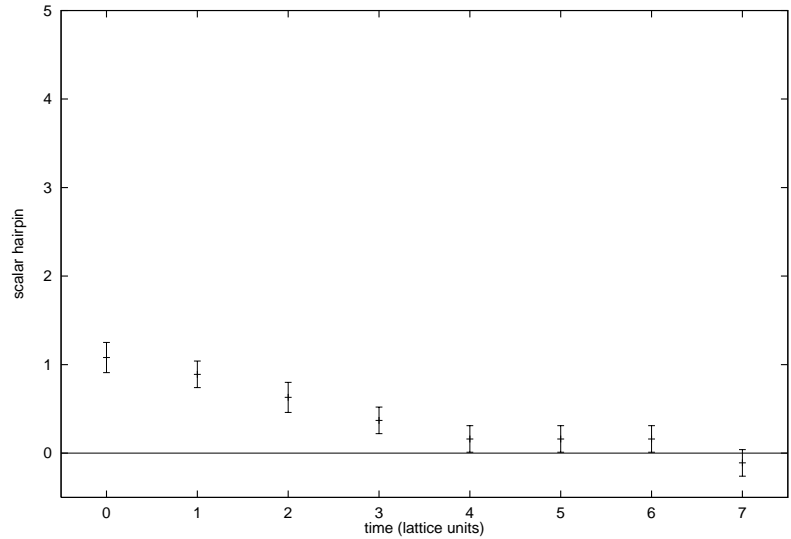}
\end{center}
\caption{The scalar double-hairpin correlator. }
\label{fig:scprop}
\end{figure}

Even without any detailed analysis, the overall empirical OZI pattern of Table
\ref{tab:OZI} is strikingly confirmed by the lattice results. This is
easily seen
from the size of the various double-hairpin correlators. In Figs.
\ref{fig:psprop}-\ref{fig:scprop}, we have plotted
the double-hairpin correlators for the pseudoscalar, vector, axial vector,
and scalar sources. All plots have the same scale for comparison. The
calculations have been done for 9 different choices of quark
mass. The data shown in the figures
are from one of the lightest quark masses, for which the  pion mass is about
300 MeV ($m_\pi a = 0.266 \pm 0.004$). The results quoted in 
Table \ref{tab:latticeOZI} are chirally
extrapolated to the physical pion mass. 
The errors in Figs. 6-9 and in Table \ref{tab:latticeOZI} 
are statistical only. By far the largest and longest-range
correlator is the pseudoscalar correlator of Fig. \ref{fig:psprop}. This is
expected for two reasons: the anomaly introduces a large double-hairpin vertex
responsible for the large
$\eta '$ mass, and, as explained above, in the quenched approximation the
external
$\bar{q}q$ meson propagators on either side of the double-hairpin vertex are
light Goldstone bosons. The results extracted from Fig. \ref{fig:psprop} have
been reported in Ref
\cite{chlogs}.

   Compared to the very strong pseudoscalar double-hairpin, the vector and
axial vector
double-hairpins of Figs. \ref{fig:vecprop} and \ref{fig:avecprop} are
dramatically suppressed,
consistent with the empirical observations described in Section
\ref{sec:background}. Since quenched lattice QCD gives reasonable values
for the
three-point functions associated with the meson virtual loop processes
depicted in
Fig.
\ref{fig:OZIloop}, these results provide not only a first derivation of the OZI
rule from QCD, but also a dramatic example of the evasion in QCD of the
``second
order paradox" described in Section \ref{sec:background} and a confirmation
of the
fact that in a complete calculation a conspiracy of the type described in
Section
\ref{sec:resolution} must occur. (Of course the results reported here
include not only the meson loop contributions but also the other time
orderings of the  double-hairpin graphs of Fig.
\ref{fig:OZI}(b).) We in fact see no significant signals in the vector and
axial vector channels and so
report in Table
\ref{tab:latticeOZI} only  one standard deviation
upper bounds.

   As described in Section \ref{sec:resolution} and illustrated in Fig.
\ref{fig:closure}, if the conspiratorial cancellation amongst meson loops is
associated with $^3P_0$ pair creation, one would expect $A_{OZI}^{0^{++}}$
to be very large. Fig.
\ref{fig:scprop} shows this behaviour: after
taking into account the heavier mass of the scalar meson
(about 1.3 in lattice units \cite{a0}),
we find that the scalar OZI amplitude is comparable in size to the pseudoscalar
amplitude  but of the opposite sign (see Table \ref{tab:latticeOZI}). A full
amplitude ${ A}_{OZI}$ in general has glueball, instantaneous, and loop
contributions, and in a given amplitude, any or all of these components might
be important. (Recall, for example,  that
while the loop contribution to ${ A}_{OZI}^{0^{-+}}$ is believed to be
small
\cite{missingpseudoscalar}, the full ${ A}_{OZI}^{0^{-+}}$ is
large.)  That the
measured
$A_{OZI}^{0^{++}}$
is actually consistent in sign and magnitude with the hadronic loop
contribution predicted by the quark model has interesting
implications which we will discuss below. A large and negative ${ A}_{OZI}^{0^{++}}$  has been
previously reported in Ref. \cite{MichaelScalar}.

To obtain the quantitative results for the OZI mixing amplitudes quoted in
Table
\ref{tab:latticeOZI}, we carried out an analysis similar to that used to
obtain the
$\eta'$ mass from the pseudoscalar double-hairpin \cite{lat98,chlogs}. For
that case, the
time-dependence of the pseudoscalar double-hairpin correlator corresponding
to Fig.
\ref{fig:OZI}(b) was found to be quite well described by a ``double-pole'' form
consisting of a
$p^2$-independent double-hairpin insertion  between a pair of meson
propagators (see also
Ref. \cite{mass2}). In momentum space
\begin{equation}
\tilde{\Delta}_h(p) = - ~ f_P\frac{1}{p^2+m_{\pi}^2} {\bf A}_{OZI}^{0^{-+}}
\frac{1}{p^2+m_{\pi}^2}f_P
\end{equation}
where $f_P$ is the vacuum-to-one-particle matrix element
\begin{equation}
f_P = \langle 0|\bar{q} i \gamma^5q|\pi(p)\rangle
\end{equation}
and ${\bf A}_{OZI}$ is the (mass)$^2$ version of the $A_{OZI}$ defined
previously
\cite{mass2} (called $m^2_0$ in Refs. \cite{lat98,chlogs}). This gives a
time-dependent double-hairpin correlator at zero 3-momentum of the form
\begin{equation}
\label{eq:fit}
\Delta_h({\bf p}=0; t) = - ~\frac{f_P^2
{\bf A}_{OZI}^{0^{-+}}}{4m^3_{\pi}}(1+m_{\pi}t)e^{-m_{\pi}t} +  (t\rightarrow
(Na-t))
\label{eq:hairpinfit}
\end{equation}
to be compared to the usual valence quark (e.g., isovector) correlator
corresponding to Fig. \ref{fig:OZI}(a)
\begin{equation}
\tilde{\Delta}_v(p) = f_P\frac{1}{p^2+m_{\pi}^2}f_P
\end{equation}
which gives
\begin{equation}
\Delta_v({\bf p}=0; t) = \frac{f_P^2}
{2m_{\pi}}e^{-m_{\pi}t} +  (t\rightarrow
(Na-t))~~.
\label{eq:valencefit}
\end{equation}
(The relative sign of Eqs. (\ref{eq:hairpinfit}) and (\ref{eq:valencefit}) is
tricky; with our convention a positive ${\bf A}_{OZI}$ makes a
positive contribution to the (mass)$^2$ of a state.) ~~~ Since the values of
$f_P$ and 
$m_{\pi}$ can be 

\bigskip

\noindent separately determined from fitting
Eq. (\ref{eq:valencefit}) to the valence quark correlator, the double-hairpin
vertex insertion ${\bf A}_{OZI}^{0^{-+}}$ can be  determined by a
one-parameter fit of (\ref{eq:fit}) to the overall size of the double-hairpin
correlator. A similar analysis
of the scalar double-hairpin led to the result quoted in Table
\ref{tab:latticeOZI}, while  for the other channels such analyses provided
the quoted  upper bounds for
the very tiny mixing amplitudes in these channels.

\section {Conclusions}
\label{sec:conclusions}

   The most straightforward conclusions of this work are
that QCD can explain the OZI rule in channels where it is observed and
that it predicts that $A_{OZI}^{0^{++}}$ is large and negative 
\cite{MichaelScalar}.  This supports the
quark model's  explanation of the dynamical suppression of the typical scale of
hadron-loop-induced OZI violation below
$1/N_c$ expectations, and in so doing provides further evidence for the
standard $^3P_0$ pair creation amplitude, since this is the critical feature
which produces this result.

     While the precise consequences are unclear, the implications for
phenomenology are serious.  With $A_{OZI}^{0^{++}}$ large, the lightest
scalar meson nonet
(the $1P$ states) will be close to the $SU(3)$ limit.  We may therefore
expect an octet of scalar mesons in the 1400 MeV range with the other $1P$
states while the nearly
singlet scalar state will be substantially lower in mass.  Thus the usual
assumption of phenomenological analyses that
this region will contain the unmixed nonet of $1P$ states and the scalar
glueball
is incorrect. For example, this region might well contain the isoscalar state
of the  $2P$ nonet. In addition, since
$A_{OZI}^{0^{++}}$ is comparable to the
$1P-2P$
splitting, there is no reason to assume that either the $1P$ or $2P$
singlet's properties can be related by nonet symmetry to those of its octet.
The net
effect is that the definitive extraction of the glueball state from the scalar
meson spectrum may be quite difficult.

     Given the importance of this task, it is certainly worthwhile
to study the scalar mesons more carefully in the light of this result
\cite{LeeWeingarten}. On the lattice it might be possible to obtain the matrix
of OZI-violating amplitudes connecting the $\omega$-like and $\phi$-like $1P$
and $2P$ states; in models the low-lying scalar meson spectrum can be studied
including the effects of a strong annihilation channel.

     Perhaps most critical is to use quenched
lattice calculations of the mixed propagators from quarkonia to
glueballs \cite{LeeWeingarten} to help resolve the scalar meson OZI violation
reported here into the contributions of $q \bar q' q' \bar q$
intermediate states, purely gluonic intermediate states associated with
``true double-hairpin" graphs, and instantaneous
contributions.
Ultimately, quenched lattice calculations of three-point functions could
directly check the predicted negative loop contributions to $A_{OZI}^{0^{++}}$
by measuring the vertex functions which are the ``raw ingredients" of the quark
model calculation.  In particular, in other than the
$0^{++}$
channel, one should see the required magnitudes and {\it opposite signs}
of the virtual $P$-wave decays to two $\ell=0$ mesons and the $S$-wave
decays to
one $\ell=1$
and one $\ell=0$ meson required to build up the near cancellation that is at
the heart of the quark model mechanism.  In contrast, for $0^{++}$ mesons
these channels should have the same sign.

\section {Discussion}
\label{sec:remarks}

   The results described here clearly have serious implications for
the spectroscopy of $0^{++}$ states, and define the series of investigations
described above required
to clarify the physics behind $A_{OZI}^{0^{++}}$.  Such investigations are
not only
important for their impact on phenomenology, however.  They are also
important because our results highlight other more fundamental questions 
raised long ago by Witten \cite{WittenUA(1)},  on the apparent conflict
between the instanton solution of the $\eta'$ mass (i.e., $U_A(1)$) problem
and
the large $N_c$ limit.  The quark model mechanism for the loop
contributions to $A_{OZI}^{0^{++}}$ is based on large $N_c$.
While our discussion of the $^3P_0$ model has focused on its prescription
for the quantum numbers
of the created $q \bar q$ pair, it is also an essential ingredient of the
model that this pair
creates $(q \bar q')+(q' \bar q)$ and not
$(q \bar q)+(q' \bar q')$ mesons, i.e., that it
respects the OZI rule at tree level.  The physical picture behind this feature
of the model is that pair creation (at order $1/N_c$) occurs by the
breaking of the color
flux tube connecting $q$ and $\bar q$.  More
generally, as mentioned
above, this limit provides the only known field-theoretic basis for
the success of not only the valence quark model, but also of Regge
phenomenology, the narrow
resonance approximation, and  many of the systematics
of hadronic spectra and matrix elements
\cite{LargeNc,DTE,ManoharJenkins,Lebed}.
In contrast, it is
widely believed that the $U_A(1)$ problem is solved through instanton
contributions to the axial anomaly.  However, as emphasized by Witten,
instantons vanish like $e^{-N_c}$ and so do not appear in the large $N_c$
expansion.  ``Insofar as [instantons play] a significant role in the
strong interactions, the large $N_c$ expansion must be bad.  It is
necessary to choose between the two." \cite{Wittenquote}
~Note that these arguments draw an important distinction between 
semiclassically calculated instanton effects, which vanish like
$e^{-N_c}$, and more general topological gauge fluctuations, which
{\it can} contribute at order $1/N_c$ to $m_{\eta '}$. The real issue is not
whether there are large fluctuations of $F \tilde F$ in the QCD vacuum, but
whether these fluctuations arise as local semiclassical lumps with quantized
winding numbers or simply as a result of the generically large gauge
fluctuations of a confining vacuum.

     To place this conflict in context, recall Eqs. (1) and (2).  From
Section \ref{sec:lattice} it is apparent that the amplitude for any of $N_f$
massless
$q \bar q$ pairs to annihilate to any other pair is the same, i.e.,
that $\Delta T$ does indeed have the form of the $N_f=2$ matrix shown in
Eq. (2).
As explained earlier, this is consistent with the 't Hooft instanton
interaction since the ``scattering" amplitude $S$ in Eq. (1)
contains a contribution $-A$ from instantons.  Thus to leading order
in $A$ the decomposition of Eqs. (1) and (2) is general and the
analyses of OZI violation in Refs. \cite{DGG,NImix,FritzschMinkowski} - - -
including that in the pseudoscalar sector - - -  are valid.  It follows
that from a purely phenomenological perspective it is irrelevant whether
or not there is an instanton contribution to hadronic physics: a
phenomenology with $A_{OZI}^{0^{-+}}\neq 0$ is ``legal" in any case,
since the anomaly allows a resolution of the $U_A(1)$ problem with
\cite{instantons,tHooftInstantons} or without \cite{WittenUA(1),Veneziano}
instantons.  What remains unclear is the physics behind the
annihilation amplitudes.  Since
a lattice simulation sums over all paths, it contains the instantons
as tunnelling events between classical vacua, but the Feynman diagrams of
QCD, which represent the quantum corrections around these vacua, are
incapable of representing  instanton physics.  Thus if
instantons are
important in QCD, Feynman diagrams would have to be supplemented by
effective interactions (like the 't Hooft interaction).  As noted by
Witten \cite{WittenUA(1)}, the foremost victim of the failure of
Feynman diagrams implied if instantons are important would be the large
$N_c$ expansion, since it assumes that all-orders properties of the QCD
Feynman diagrammatic expansion are properties of QCD.

     The observations reported in this paper on $A_{OZI}^{0^{++}}$ add one
more item to a
growing and closely linked set of issues where the physics of instantons
and the physics of large $N_c$ confront each other.  Assuming that
confinement and the Nambu-Goldstone mechanism \cite{NambuGoldstone} are
properties of the all-orders Feynman diagrammatic expansion of QCD,
the large $N_c$ expansion provides a consistent framework
embracing all strong interaction phenomena.  Among these phenomena are
the hadron spectrum for all flavors of hadrons (including the
$1/N_c$-suppressed hadronic widths which seem to be critical to
$A_{OZI}^{0^{++}}$),
the OZI rule (now including $A_{OZI}^{0^{++}}$), and the $q \bar q$
condensate.  As Witten argued
long ago \cite{WittenUA(1)}, given the $U_A(1)$ anomaly and confinement,
the large
$N_c$ limit is also capable of explaining the $\eta'$ mass at order
$1/N_c$ {\it without} instantons.

     While its limited range of applicability makes it somewhat less
attractive for phenomenology (instantons offer a competing explanation only
for the
properties of the lightest $SU(3)_f$ hadrons),the instanton picture 
\cite{ShuryakReview} has received
strong support from recent lattice results \cite{Negele}. Measurements
of the topological charge \cite{WittenUA(1)} of
``cooled" gauge configurations show that in such circumstances this charge is
quantized and localized as expected for
instantons.  Moreover, the zero-modes of the Dirac operator associated with
the solution of the
$U_A(1)$ problem and the near-zero-modes associated with the $q \bar q$
condensate
are also localized and in ``cooled" configurations can be associated with
these same instantons.  The lattice results on these and other hadronic
properties are consistent with the instanton liquid model \cite{ShuryakReview}.

     Since, as argued by Witten, confinement can replace instantons as the
source
of the $U_A(1)$ anomaly and since confinement can also produce a space-time
localization of the origin
of the $\eta'$ mass and of the $q \bar q$ condensate, in our view the true
origin of these effects remains
unsettled.  The results of this paper may help to resolve this situation since
for $A_{OZI}^{0^{++}}$ the two competing pictures
lead to  mechanisms that are very distinct.
Flux-tube-breaking pair creation, a prototypical large $N_c$ phenomenon, led
to the prediction that the hadron loop contribution to $A_{OZI}^{0^{++}}$ is
large and negative as found here.
Moreover, as stated in the beginning of this paper, quark models, with
their confined constituent quarks, naturally generate a large positive
$A_{OZI}^{0^{-+}}$
\cite{DonoghueGomm}.  In this case the loop contribution should be
typically small \cite{missingpseudoscalar}, and the large positive quark
model amplitude is associated with an instantaneous interaction. Instantons,
through the instantaneous 't Hooft interaction, would lead to  a
superficially similar pattern of OZI violation:  a large positive
$A_{OZI}^{0^{-+}}$ and a large negative
$A_{OZI}^{0^{++}}$. However, the origins of the large negative
$A_{OZI}^{0^{++}}$ are very different in the two cases: the instanton
$A_{OZI}^{0^{++}}$ is associated with an instantaneous contribution while the
quantitative similarity between the quark model prediction and our measured
$A_{OZI}^{0^{++}}$ suggests that this amplitude is associated instead with
the meson loop contributions. 

     Our result thus favors the large $N_c$ and not the instanton
interpretation of the solution to the $\eta '$ mass problem. Nevertheless,
while suggestive, the quark model prediction is not of sufficient
quantitative accuracy for this conclusion to be reliable. Fortunately, with
recent advances in lattice methods and in  computing power, we believe
that the results we have described here can not only be improved but also
understood more deeply. In particular,  through the program we described of
decomposing the OZI-violating amplitudes into their component parts, it
should be possible to define the mechanism driving $A_{OZI}^{0^{++}}$. We
also believe it will be particularly fruitful to define and test
confinement-based interpretations of the lattice results on such quantities
as the topological susceptibility, the localization of zero modes, the
correlation function of the topological charge operator, and the space-time
association of the $q \bar q$ condensate with the topological charge. Through
such studies, the conflict between large
$N_c$ and instanton physics can at last be resolved.

\vfill\eject

{\centerline {\bf ACKNOWLEDGEMENTS}}

\medskip

    We are grateful to Stephen Sharpe and Thomas Schaefer and to Chris
Michael for alerting us to a serious sign error in the first version of
this paper and to important references which had escaped our attention. This
work was supported by DOE contract DE-AC05-84ER40150 under which the
Southeastern Universities Research Association (SURA) operates the Thomas
Jefferson National Accelerator Facility. The work of H.B. Thacker was
supported in part by the Department of Energy under grant DE-FG02-97ER41027.

\bigskip

{

\end{document}
\centerline {\bf REFERENCES}}

\begin{references}


\bibitem{Zweig}
G. Zweig, CERN Report No. 8419 TH 412, 1964 (unpublished);
reprinted in {\it Developments in the Quark Theory of Hadrons},
edited by D. B. Lichtenberg and S. P. Rosen
(Hadronic Press, Massachusetts, 1980). The history of the discovery of the
quark model (or ``aces")
as seen by Zweig is related in  ``Baryon 1980", Proceedings of the IVth
International
Conference on Baryon Resonances, ed. N. Isgur (Toronto, 1980), p. 439.


\bibitem{otherOZI}
S. Okubo, Phys. Lett. {\bf 5}, 165 (1963);
Phys. Rev. D {\bf16}, 2336 (1977);
J. Iizuka, K. Okada, and O. Shito, Prog. Th. Phys. {\bf35},
1061 (1966); J. Iizuka, Prog. Th. Phys. Suppl. {\bf37-38}, 21 (1966).

\bibitem{DGG} A. De R\'ujula, H. Georgi, and S. L. Glashow, Phys. Rev.
D{\bf 12}, 147 (1975).


\bibitem{NImix} N. Isgur, Phys. Rev. {\bf D12}, 3770 (1975); {\bf D13}, 122
(1976).


\bibitem{FritzschMinkowski} H. Fritzsch and P. Minkowski, Nuovo Cimento
{\bf 30A}, 393 (1975).

\bibitem{subtlety} There is an interesting subtlety associated with this
decomposition. Fig. \ref{fig:OZI}(b) includes time-orderings in which the pair
creation occurs before the annihilation (see Fig. \ref{fig:OZIloop})
corresponding to
meson loop processes. However, the complete set of meson loop processes
produce not
only Fig. \ref{fig:OZI}(b) but also Z-graphs of Fig. \ref{fig:OZI}(a) and
graphs
with closed $q \bar q$ loops inserted into Fig. \ref{fig:OZI}(a). Thus a
consistent
treatment actually requires that $S$ be not just the simple valence graph
shown, but
also include graphs with internal $q \bar q$ loops \cite{NImix}.

\bibitem{UA(1)Problem} M. Gell-Mann, R.J. Oakes, and B. Renner, Phys. Rev.
{\bf 175}, 2195 (1968);
S. Weinberg, Phys. Rev. {\bf D12}, 3583 (1975).


\bibitem{NambuGoldstone} Y. Nambu, Phys. Rev. Lett. {\bf 4}, 380 (1960); J.
Goldstone, Nuovo Cim. {\bf 19}, 154 (1961).



\bibitem{instantons} A.M. Polyakov, Phys. Lett. {\bf 59B}, 82 (1975);
Nucl. Phys. {\bf B121}, 429 (1977);
A.A. Belavin, A.M. Polyakov, A. Schwartz, and Y. Tyupkin, Phys. Lett. {\bf
59B}, 85 (1975);
C. Callan, R. Dashen, and D.J. Gross, Phys. Lett. {\bf 63B}, 334 (1976);
R. Jackiw and C. Rebbi, Phys. Rev. Lett. {\bf 37}, 172 (1976).

\bibitem{tHooftInstantons}
G. 't Hooft, Phys. Rev.  Lett. {\bf 37}, 8 (1976);
Phys. Rev. {\bf D14}, 3432 (1976).


\bibitem{WittenUA(1)} E. Witten, Nucl. Phys. {\bf B149}, 285 (1979); {\bf
256}, 269 (1979).

\bibitem{Veneziano} G. Veneziano, Nucl. Phys. B{\bf 159}, 213 (1979).

\bibitem{LargeNc}
G. 't Hooft, Nucl. Phys. {\bf B72}, 461 (1974);
E. Witten, Nucl. Phys. {\bf B160}, 57 (1979).


\bibitem{mass2} The extraction of the OZI amplitudes from the data can
be complex. If
$A_{OZI}$ is small,
$\omega-\phi$-like mixing is small as is mixing to excited nonets and
glueballs, so
$A_{OZI}$ may readily be extracted from the relation
$m_{I=0}=m_{I=1}+2 A_{OZI}$. Note that since this extraction is in the
$SU(2)_f$
sector, issues of $SU(3)$-breaking do not arise. Also note that  the
relation between $A_{OZI}$ and ${\bf A}_{OZI}$ defined by the analogous
(mass)$^2$ formula $m^2_{I=0}=m^2_{I=1}+2 {\bf A}_{OZI}$
is simply
$A_{OZI}={\bf A}_{OZI} /2 m_{I=1}$
at the light quark mass scale.  When
$A_{OZI}$ is large, mixing  and $SU(3)$ breaking can become
important. In such circumstances, no simple
comparison to the data can be made without further assumptions. Since excited
nonets will normally be 0.5 to 1.0 GeV away, when $A_{OZI}$ is as large as
it is
in the pseudoscalars and scalars, the validity of ignoring internonet mixing is
certainly questionable. If one nevertheless makes this assumption, then in the
$SU(3)$ limit one would have $m_0=m_{I=1}+3 A_{OZI}$ or alternatively
$m^2_0=m^2_{I=1}+3 {\bf A}_{OZI}$ where $m_0$ is the mass of the $SU(3)$
singlet
meson, leading to the relation
$A_{OZI}={\bf A}_{OZI} / [m_0+m_{I=1}] $. For weak $SU(3)$ breaking these
formulas
simply become
$m'_0=2m_{I={1 \over 2}}-m'_{I=0}+3 A_{OZI}$ and $m'^2_0=2m^2_{I={1 \over
2}}-m'^2_{I=0}+3 {\bf A}_{OZI}$ where now $m'_0$ and $m'_{I=0}$ are the
masses of
the mainly singlet and mainly octet mesons (e.g., the $\eta '$ and the $\eta$,
respectively). It is this latter formula that is commonly used in defining
${\bf
A}_{OZI}^{0^{-+}}$ in the Witten-Veneziano formula
\cite{WittenUA(1),Veneziano} and
which leads to $3{\bf A}_{OZI}^{0^{-+}}=$(0.85 GeV)$^2$. These broken
$SU(3)$ relations lead to
$A_{OZI}=\Bigl[{ 
{ {m'_0+m'_{I=0} \over 2}- m_{I={1 \over 2}} }
\over
{ {m'^2_0+m'^2_{I=0} \over 2}- m^2_{I={1 \over 2}} } 
}\Bigr]  
{\bf A}_{OZI}$.
The inaccuracy of these formulas from both internonet mixing and higher order
corrections in
$SU(3)$-breaking are not obviously small. Thus, given the dynamical
assumptions
required to relate a large value of
$A_{OZI}$ to observed masses, the empirical value of $A^{0^{-+}}_{OZI}$ (or
${\bf
A}^{0^{-+}}_{OZI}$) is quite uncertain (as is $A^{0^{++}}_{OZI}$ and ${\bf
A}^{0^{++}}_{OZI}$) and therefore only semiquantitative statements about
these amplitudes can be made at this time. To convert the lattice
propagator data for ${\bf A}_{OZI}$ to $A_{OZI}$ in Table
\ref{tab:latticeOZI} for comparison to the $A_{OZI}$ quoted in Table
\ref{tab:OZI}, we note that in the single flavor case considered, if we
once again make the assumption of small mixing with other states, the
mass and (mass)$^2$ formulas are $m=m_{I=1}+A_{OZI}$ and
$m^2=m^2_{I=1}+{\bf A}_{OZI}$. (Note that we have identified the mass
without the annihilation amplitude as being the $I=1$ mass that would be
found for two or more flavors.) It follows that $A_{OZI}=
\sqrt{m^2_{I=1}+{\bf A}_{OZI}}-m_{I=1}$. For small $A_{OZI}$ this
relation gives $A_{OZI} \simeq {\bf A}_{OZI} /2 m_{I=1}$ and (as in our
extraction of these quantities from the data) in this case our assumption
of small mixing is justified. For large $A_{OZI}$ such an assumption is
probably not good; however, the lattice relation can be systematically
improved by measuring additional elements of the mass  and
(mass)$^2$ matrices should such an improvement be justified by an
improvement in the relevant experimental data.



\bibitem{LipkinOZI}
H.J. Lipkin, Nucl. Phys. {\bf B291}, 720 (1987);
Phys. Lett. {\bf B179}, 278 (1986);
Nucl. Phys. {\bf B244}, 147 (1984);
Phys. Lett. {\bf B124}, 509 (1983).

\bibitem{caveat} There is a loophole in this argument. It is possible that
instead
there could be a conspiracy between the hadronic loop processes and other
sources of
OZI-violation.


\bibitem{GIonOZI}
P. Geiger and N. Isgur, Phys. Rev. D {\bf44}, 799 (1991);
Phys. Rev. Lett. {\bf67}, 1066 (1991);
Phys. Rev. D {\bf47}, 5050 (1993);
P. Geiger, {\it ibid.} {\bf49}, 6003 (1993).


\bibitem{3P0}
L. Micu, Nucl. Phys. {\bf B10}, 521 (1969);
A. Le Yaouanc, L. Oliver, O. Pene, and J.-C.
Raynal, Phys. Rev. D {\bf8}, 2233 (1973); Phys. Lett. {\bf B71  }, 397
(1977); {\it ibid.} {\bf B72 }, 57 (1977);
W. Roberts and B. Silvestre-Brac, Few Body Syst. {\bf 11}, 171 (1992);
P. Geiger and E.S. Swanson, Phys. Rev. D {\bf50}, 6855 (1994);
R. Kokoski and N. Isgur, Phys. Rev. D {\bf35},
907 (1987);
Fl. Stancu and P. Stassart,
Phys. Rev. D {\bf38}, 233 (1988); {\bf39}, 343 (1989);
{\bf41}, 916 (1990); {\bf42}, 1521 (1990);
S. Capstick and W. Roberts, Phys. Rev. D {\bf47}, 1994 (1993);
Phys. Rev. D {\bf49}, 4570 (1994);   
E.S. Ackleh, T. Barnes, and E.S. Swanson, Phys. Rev. D {\bf54}, 6811
(1996); and for a recent and detailed description of the implications
of the $^3P_0$ model for mesons, see T. Barnes, F.E. Close, P.R. Page, and
E.S. Swanson, Phys. Rev. D {\bf55}, 4157 (1997).



\bibitem{ReggeExotics} C. Schmid, D.M. Webber, and C. Sorensen, Nucl. Phys.
{\bf B111}, 317 (1976);
E.L. Berger and C. Sorensen, Phys. Lett. {\bf 62B}, 303 (1976).

\bibitem{missingpseudoscalar} Unfortunately, Ref. \cite{GIonOZI} did not
check the
loop contribution to $A_{OZI}^{0^{-+}}$ since the $U_A(1)$ anomaly
guaranteed that
OZI violation in this sector would be large so that there was no second order
paradox to evade. However, given the identified mechanism of the
suppression of the
loop diagrams, there is every reason to suppose that the loop contribution to
$A_{OZI}^{0^{-+}}$ is as small as in any other nonet with $J^{PC} \neq 0^{++}$.

\bibitem{Kuramashi} Y. Kuramashi, M. Fukugita, H. Mino, M. Okawa, and A. Ukawa,
Phys. Rev. Lett. {\bf 72},  3448 (1994).

\bibitem{lat98} W. Bardeen, A. Duncan, E. Eichten, and H. Thacker, Nucl. Phys.
B (Proc. Suppl.) {\bf 73},  243 (1999).

\bibitem{chlogs} W. Bardeen, A. Duncan, E. Eichten, and H. Thacker, Nucl.
Phys B
(Proc. Suppl.) {\bf 83-84},  215 (2000).

\bibitem{scales} A.X. El-Khadra, G.M. Hockney, A.S. Kronfeld, and P.B. Mackenzie,
Phys. Rev. Lett. {\bf 69},  729 (1992).

\bibitem{a0} We have determined the $\rho$,  $a_0$, and $a_1$ masses from
chirally extrapolated fits to our data on the valence (i.e., isovector)
propagators associated with Fig.
\ref{fig:OZI}(a),
which give 
$m_{\rho} \simeq 0.8$ GeV,
$m_{a_1} \simeq 1.4$ GeV, and
$m_{a_0} \simeq 1.6$ GeV. The latter two masses seem to be too high. It is
quite possible that they are especially sensitive to quenching since both the
$a_1$ and the $a_0$ have (or are predicted to have)  very strong S-wave
couplings to nearby thresholds.  (See, for example, R. Kokoski and N. Isgur,
Phys. Rev. D {\bf35}, 907 (1987); N. Isgur, Phys. Rev. D {\bf57}, 4041
(1998).) We have found no other recent lattice measurements of these masses
close to the chiral limit (see, however, Ref. \cite{MichaelScalar}), or of
other spectroscopic properties of the P-wave mesons, and consequently hope to
report on their spectroscopy in a forthcoming publication. In any event, at
the level of accuracy relevant to the results reported here, these $10\%$
effects are not significant.

\bibitem{MichaelScalar} C. Michael, M. S. Foster and C. McNeile (UKQCD
Collaboration), Nucl. Phys. B (Proc. Suppl.) {\bf 83-84}, 185  (2000).


\bibitem{LeeWeingarten} W. Lee and D. Weingarten, Nucl. Phys. B (Proc.
Suppl.) {\bf 63}, 194  (1998); Nucl. Phys. B (Proc. Suppl.) {\bf 73},
249  (1999); Phys. Rev. D {\bf61}:014015 (2000). See also Ref.
\cite{MichaelScalar} on $^3P_0$-glueball mixing.

\bibitem{DTE} The Dual Topological Expansion is very closely related to the
large
$N_c$ expansion. See G.F. Chew and C. Rosenzweig, Nucl. Phys. {\bf B}104, 290
(1976); G. Veneziano, Proc. $9^{th}$ Ecole d'Et\'e de Physique des Particules,
Gif-sur-Yvette, 1977 (1978), vol. 2, p. 23.

\bibitem{ManoharJenkins} R. Dashen and A. Manohar, Phys. Lett. {\bf B315},
425 (1993);
{\bf B315}, 438 (1993);
R. Dashen, E. Jenkins and A. Manohar, Phys. Rev. D{\bf49}, 4713 (1994);
D{\bf51}, 2489 (1995).

\bibitem{Lebed} Elizabeth Jenkins and Richard F. Lebed, Phys. Rev. D {\bf
52}, 282 (1995).

\bibitem{Wittenquote} E. Witten,  Nucl. Phys. {\bf B149}, 285 (1979) on
page 286.


\bibitem{ShuryakReview} For a review, see T. Sch\"afer and E. Shuryak, Rev.
Mod. Phys. {\bf 70},
323 (1998).

\bibitem{Negele} For a review, see J. Negele,
Nucl. Phys. {\bf B} (Proc. Suppl.) {\bf 73}, 92 (1999).


\bibitem{DonoghueGomm} John F. Donoghue and Harald Gomm, Phys. Rev. D {\bf
28}, 2800 (1983).




\end{references}
